\pdfoutput=1
\documentclass[twocolumn,longbibliography,superscriptaddress,floatfix,prl]{revtex4-1}

\usepackage{graphicx}
\usepackage{dcolumn}
\usepackage{mathtools, nccmath}
\usepackage{bm}
\usepackage{xcolor}
\usepackage{amsmath}
\usepackage[export]{adjustbox}
\usepackage{framed}
\usepackage[]{graphicx}
\usepackage[caption=false]{subfig}
\usepackage{hyperref}
\usepackage[mathlines]{lineno}
\usepackage{textcomp}
\usepackage{gensymb}
\usepackage[super]{nth}
\usepackage{graphicx,color}
\usepackage{hyperref}
\usepackage{babel}
\usepackage{booktabs}
\usepackage[normalem]{ulem}
\usepackage{dsfont}
\usepackage[percent]{overpic}

\usepackage{stackengine}

\newcommand{\pdag}{{\phantom{\dagger}}}

\newcommand{\pd}{{\phantom{\dagger}}}

\begin{document}

\preprint{APS/123-QED}

\title{Hawking fragmentation and Hawking attenuation in Weyl semimetals}

\author{Daniel Sabsovich}
\affiliation{Raymond and Beverly Sackler School of Physics and Astronomy, Tel-Aviv University, Tel-Aviv 69978, Israel}

\author{Paul Wunderlich}
\affiliation{Institute for Theoretical Physics and W\"urzburg-Dresden Cluster of Excellence ct.qmat, Technische Universit\"at Dresden, 01069 Dresden, Germany}
\affiliation{Institute for Theoretical Physics, Goethe University Frankfurt, Max-von-Laue-Straße 1, 60438 Frankfurt a.M., Germany}

\author{Victor Fleurov}
\affiliation{Raymond and Beverly Sackler School of Physics and Astronomy, Tel-Aviv University, Tel-Aviv 69978, Israel}

\author{Dmitry I. Pikulin}
\affiliation{Station Q, Microsoft Corporation, Santa Barbara, California 93106-6105, USA}
\affiliation{Microsoft Quantum, Redmond, Washington 98052, USA}

\author{Roni Ilan}
\affiliation{Raymond and Beverly Sackler School of Physics and Astronomy, Tel-Aviv University, Tel-Aviv 69978, Israel}

\author{Tobias Meng}
\affiliation{Institute for Theoretical Physics and W\"urzburg-Dresden Cluster of Excellence ct.qmat, Technische Universit\"at Dresden, 01069 Dresden, Germany}

\date{\today}

\begin{abstract}
We study black and white hole analogues in Weyl semimetals with inhomogenous nodal tilts. We study how the presence of a microscopic lattice, giving rise to low-energy fermion doubler states at large momenta that are not present for elementary particles, affects the analogy between Weyl Hamiltonians and general relativity. Using a microscopic tight-binding lattice model, we find the doubler states to give rise to Hawking fragmentation and Hawking attenuation of wavepackets by the analogue event horizon. These phenomena depend on an analogue Hawking temperature, and can be measured in metamaterials and solids, as we confirm by numerical simulations.
\end{abstract}

\maketitle

\emph{Introduction. } 
Relativistic semimetals are solids whose low-energy Hamiltionian takes the form of the Dirac equation \cite{Dirac1928}. These materials made fruitful connections between high-energy and solid state physics \cite{Volovik2003,Volovik2020}. Besides the two-dimensional Dirac material graphene \cite{wallace1947Graphite,Novoselov2005,Zhang2005,Castro2009}, recent studies have also focused on three-dimensional Dirac and Weyl semimetals \cite{Weyl1929,volovik1987,Burkov2011b,Xiangang2011,turner2013,Hosur2013,Armitage2018}. 

The extent to which the formal analogy between low-energy Hamiltonians and the Dirac equation implies a parallelism in experimental quantities varies. Some high-energy phenomena such as Klein tunnelling \cite{OBrien2016,Yesilyurt2016} or the chiral anomaly \cite{Nielsen1983,Fukushima2008,Zyuzin2012,Vazifeh2013} have direct analogues in solids. For other effects, such as the mixed axial-gravitational anomaly, a direct equivalent has not yet been observed. Nevertheless, even in these case, a carefully designed experiment can unearth insightful connections between high-energy physics and solids \cite{neiman_11,jensen_13,Landsteiner2016,lucas_16,Gooth2017,stone_mixed_anomal,Volovik2003}.

One of the most intriguing phenomena in general relativity are black holes \cite{Schwarzschild1916,frolov2011} and their associated Hawking radiation \cite{Hawking1974,Hawking1975}. Recently, a formal analogy between the low-energy Hamiltonian of tilted Weyl nodes \cite{Volovik2014,Yong2015,Soluyanov2015} and the spacetime metric associated with black holes, suggested that Weyl semimetals with inhomogenously tilted nodes are analogous to black holes, and that they may exhibit an analogue of Hawking radiation \cite{Huhtala2002,Volovik2016}. 


\begin{figure}[t]
\captionsetup[subfloat]{labelformat=empty}
\captionsetup[subfloat]{farskip=0pt,captionskip=-100pt}
\subfloat[]{\includegraphics[width=0.94\columnwidth]{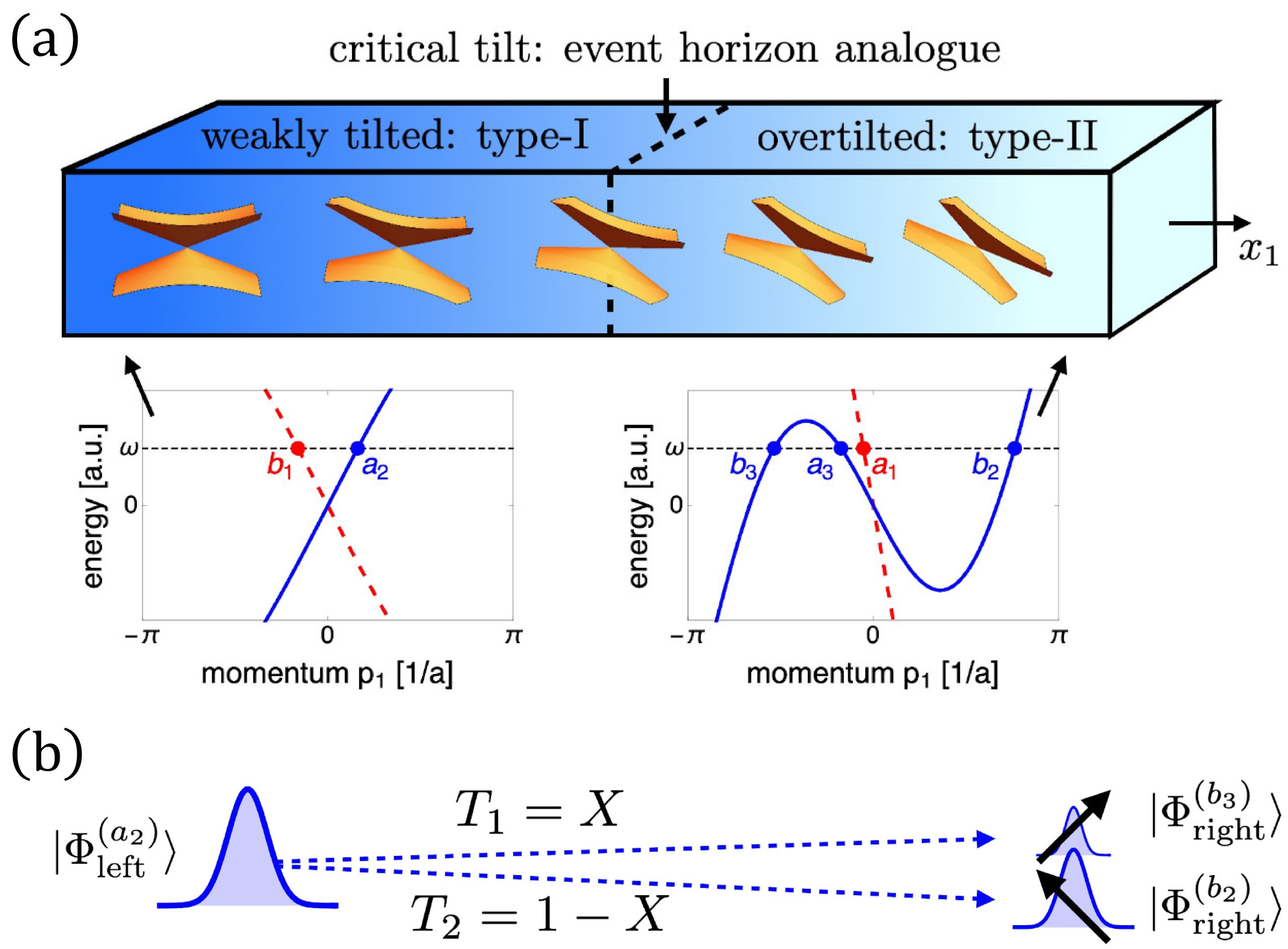}}
\vspace{-0.0in}\caption{Panel (a): sketch of a Weyl semimetal with tilt varying along $x_1$ realizing a white hole analogue. The asymptotic Hamilonians far from the central scattering region define the band structures shown below. States with a group velocity towards the horizon (incoming states) are labeled $a$, outgoing states $b$. Panel (b): An incoming wavepacket is Hawking-fragmented into outgoing wavepackets with non-parallel spins. The transmission probability $X$ depends on an effective Hawking temperature, see Eq.~\eqref{eq:hawking_expo}. Wavepackets $|\Phi_i^{(j)}\rangle$ are labeled by the state they are centered around.\label{fig:wh_system}}
\end{figure}

In the present work, we analyse how far the formal connection between (over-)tilted Weyl nodes and black holes carries, and which experimentally accessible consequences derive from it. Importantly, the formal analogy only holds for states at low energy and low momenta. The presence of an underlying lattice, however, demands the additional presence of fermionic doublers at low energies and \emph{large} momenta. These doublers are required by the one-dimensional variant of the fermion doubling theorem stating that there must be as many right-movers as left-movers in a one-dimensional lattice model \cite{Nielsen1981}. This theorem also applies along a one-dimensional momentum cut in a higher-dimensional Brillouin zone, and is not to be confused with the three-dimensional fermion doubling theorem stating that Weyl nodes appear in pairs. Our analysis shows that these doublers are in fact crucial for the discussion of Hawking-analogues in Weyl semimetals. We find that there is no direct analogue of Hawking radiation in static Weyl semimetals with inhomogenous tilts%
. In contrast, the large-momentum doublers give rise to new phenomena -- Hawking attenuation for black hole analogues and Hawking fragmentation for white hole analogues. That is, a wavepacket crossing the analogue event horizon is either attenuated by a factor determined by an effective Hawking temperature, or fragmented into pieces with probabilities set by the effective Hawking temperature. Necessarily involving doubler states, these phenomena are unique to lattice systems, and we do not expect them in cosmological black or white holes which are described by a linear spectrum.

\emph{Black holes and Hawking radiation. }
A black hole is created when an object of mass $M_{\bullet}$ is so strongly compressed that gravity forbids light to escape from its vicinity \cite{frolov2011}. The distance $|\vec{r}|=r$ from the black hole below which no escape is possible is the event horizon $r_{\text{hor}}$. Einstein's field equations \cite{einstein1915} relate the black hole to the space-time metric. For the simple case of a Schwarzschild black hole, the associated line element in Gullstrand-Painlev\'e coordinates \cite{Schwarzschild1916,painleve1921,gullstrand1922} reads
\begin{align}
ds^{2}&=c^{2}dt^{2}-\left(d\vec{r}-\vec{V}(r)\,dt\right)^{2}.\label{eq:PGMEtric}
\end{align}
The effect of the black hole is manifested in the velocity $\vec{V}(r)$ of an object free-falling toward the black hole. The event horizon $r_{\text{hor}}=2GM_{\bullet}/c^2$ depends only on the black hole's mass, the gravitational constant $G$, and the speed of light $c$. Expanding $\vec{V}$ close to the horizon yields

\begin{align}
\vec{V}(r)&=-c\sqrt{\frac{r_{h}}{r}}\,\vec{e}_r\approx-\left(c-\alpha\,(r-r_{\text{hor}})\right)\,\vec{e}_r.\label{eq:PGMEtricVelocity}
\end{align}
Here,  $\vec{e}_r$ is the radial unit vector, and $\alpha = {c^3}/{4 G M_{\bullet}}$ is related to the black hole's surface gravity $g_{\rm surf}$ (gravitational acceleration at $r_{\text{hor}}$) as $\alpha = g_{\rm surf}/c$.

Hawking showed that quantum fluctuations make black holes radiate: after the spontaneous creation of particle-antiparticle pairs close to the event horizon, one partner of the pair may escapce across the horizon. The radiation energy is provided by a reduction of the black hole's mass, which therefore evaporates. The spectrum of Hawking radiation is of black body type with the effective Hawking temperature set by \cite{Hawking1974,Hawking1975}

\begin{align}
T_{\text{Hawking}}=\frac{\hbar c^3}{8\pi G M_{\bullet} k_{\rm B}}=\frac{\hbar\,\alpha}{2\pi k_{\rm B}}.
\end{align}
The connection between the Hawking temperature and the black hole's surface gravity can be understood in two insightful ways. On the one hand, the Unruh effect \cite{Fulling1973,Davies1975,Unruh1976} states that an accelerated observer, described by the Rindler coordinates \cite{Rindler1966}, sees a thermal bath at an effective temperature that equals  $T_{\text{Hawking}}$ for a free-falling object at the event horizon (where the object is accelerated by $g_{\rm surf}=\alpha\,c$). On the other hand, a description of Hawking radiation in terms of quantum tunneling across a black hole horizon \cite{Parikh2000,Srinivasan1999} shows that the semi-classical emission rate of a particle with energy $\hbar \omega$ from a black hole is determined by $\alpha = g_{\text{surf}}/c$,

\begin{align}
\Gamma\sim e^{-\hbar \omega/k_{\rm B}T_{\text{Hawking}}}=e^{-2\pi \omega/\alpha}.\label{eq:hawking_probba}
\end{align}
For the recently detected solar mass black holes \cite{Abbott2016} and more massive black holes \cite{Graham2016}, $T_{\text{Hawking}}$ is much smaller than the cosmic microwave background temperature. This motivated Unruh \cite{Unruh1981} to propose so-called black hole analogues: systems that mimic certain aspects of black holes with much larger effective Hawking temperatures. The analogy typically does not involve spontaneous radiation, but an exponential dependence similar to Eq.~\eqref{eq:hawking_probba} of some quantity, \emph{e.g.}~transmission probabilities across the analogue of a horizon. Since Unruh's original suggestion involving space-dependent velocities of a classical fluid, similar ideas have also been pursued in optics, Bose-Einstein condensates, water waves and others \cite{Recati2009,Barcelo2011,Weinfurtner2011,Robertson2012,Vinish2016,Barcelo2019,Rosenberg2020,Aguero2020,morice_21}. Several reports on Bose-Einstein condensates and optical systems have reported the detection of signatures of quantum analogues of black hole physics \cite{Steinhauer2016,Drori2019}.

\emph{Mapping Weyl semimetals to black and white holes. }
Recently, it was  suggested \cite{Volovik2016} that Weyl semimetals with inhomogeously tilted nodes can realize both black and white hole \cite{novikov1964delayed,ne1965expansion} analogues. To appreciate this analogy, consider a Weyl Hamilonian with fixed tilt $\vec{V}_{\text{t}}$, 

\begin{align}
\mathcal{H}_{\text{Weyl}}=\pm\,v_F\,\vec{\sigma}\cdot\vec{p}+\mathds{1}\,\vec{V}_{\text{t}} \cdot\vec{p},\label{eq:tilted_weyl}
\end{align}
where $v_F>0$ is the Fermi velocity, $\vec{p}$ the three-dimensional momentum, and $\vec{\sigma}$ the vector of Pauli matrices. For small tilts, $|\vec{V}_{\text{t}}|<v_F$, the Weyl node is denoted as type-I. For large tilts  $|\vec{V}_{\text{t}}|>v_F$, the Weyl node is overtilted, and called type-II \cite{Soluyanov2015}. Put simply, the analogy between black or white holes and Weyl semimetals is a mapping between Weyl cones and lightcones, see Supplemental Material \cite{supplement}. Untilted Weyl cones correspond to lightcones in flat Minkowski spacetime \cite{minkowski1908fundamental,frolov2011}, (over-)tilted Weyl nodes to lightcones that tilt close to black or white holes. Black hole analogues correspond to the node tilting towards the horizon, white hole analogues to a tilt away from the horizon, see Fig.~\ref{fig:wh_system} (a).

This  pictorial  analogy  can be  put on firm ground via a mathematical mapping between the Weyl Hamiltonian in Eq.~\eqref{eq:tilted_weyl} and spacetime metrics \cite{Huhtala2002,Volovik2016}. The mapping relies on the definition of frame fields ${e_\mu}^i$ and their inverse ${{\underline{e}}^\nu}_j$. In general relativity, frame fields define a local orthonormal coordinate system at each point in space-time. They relate to the metric as $g_{ij} = \eta_{\mu\nu}\,{{\underline{e}}^{\mu}}_i\,{{\underline{e}}^{\nu}}_j$, where $\eta_{\mu\nu}$ is the Minkowski metric. In Weyl Hamiltonians, frame fields can formally be defined from the tensor connecting Pauli matrices and momenta. That is, a general Weyl Hamiltonian takes the form
\begin{align}
\mathcal{H}_{\text{Weyl}}^{\text{gen}}= -i\, v_F\,\sigma^\mu\,{e_\mu}^j\,\left(\partial_j+\frac{1}{2}\,{{\underline{e}}^\nu}_j\,(\partial_k\,{e_\nu}^k)\right),\label{eq:node_ti;t_framefield}
\end{align}
where $\sigma^0=\mathds{1}$, Roman indices go over the spatial coordinates, Greek coordinates go over space-time coordinates, and summation over identical indices is understood \cite{curved_spacetimes_ojanen_17}. The second term can equivalently be understood as deriving from the constraint that the Hamiltonian must be Hermitian, or as resulting from evaluating the spin connection for our effective metric. The essence of the mapping between Weyl Hamiltonians and black holes is that the frame fields ${e_\mu}^i$ defined from Eq.~\eqref{eq:node_ti;t_framefield} equal the ones of the Gullstrand-Painlev\'e-Schwarzschild metric associated with Eq.~\eqref{eq:PGMEtric}, with the tilt $\vec{V}_{\text{t}}/v_F$ taking the role of the velocity $\vec{V}/c$. In this way, a Weyl Hamiltonian in which the tilt changes continuously from $\vec{V}_{\text{t}}(r)=0$ 
to $|\vec{V}_{\text{t}}(r)|>v_F$, see Fig.~\ref{fig:wh_system} (a), can be mapped to the metric of a black or white hole in Gullstrand-Painlev\'e coordinates. The location of the critical tilt, $\vec{V}_{\text{t}}(r)=v_F$ is the analogue of the event horizon. 

\emph{Weyl semimetal event horizon analogues, and the role of fermion doublers. }
In the remainder, we study a simple lattice Hamiltonian allowing tilted Weyl nodes at low energies,
$H = \sum_{\vec{p}}\Psi_{\vec{p}}^\dagger\,\mathcal{H}\,\Psi_{\vec{p}}^\pd$, where $\Psi_{\vec{p}}^\dagger = (c_{\uparrow}^\dagger,c_{\downarrow}^\dagger)$ combines the annihilation operators for electrons of spin $\uparrow,\downarrow$ and momentum $\vec{p}$, and (using units of $\hbar=c=1$)
\begin{align}
\mathcal{H}(\vec{p})=&-t\,\sum_{i=1}^2\sin(p_ia)\,\sigma^i+t\,\left[2-\sum_{i=1}^3\cos(p_ia)\right]\sigma^3\nonumber\\ 
&- V_{\text{t},1}\sin(p_1a)\sigma^0.\label{eq:3DTBModel}
\end{align}
Here, $t$ is the nearest-neighbor hopping and $a$ the lattice spacing (we use $t=a=1$ for simplicity). 

This model exhibits two Weyl nodes at $p_1=p_2=0$ and $p_3=\pm \pi/2a$. Finite  and constant $V_{\text{t},1}$ uniformly tilts the nodes along the $p_1$-direction. Expanding $\mathcal{H}(\vec{p})$ for momenta close to the Weyl nodes yields the Hamiltonian \eqref{eq:tilted_weyl}. This low-momentum expansion faithfully describes the low-energy physics in the type-I regime of weakly tilted nodes. For the overtilted type-II regime, however, the 1+1-dimensional fermion doubling theorem \cite{Nielsen1981} demands the presence of additional fermionic doubler at momenta away from the nodes. To appreciate this point, consider a cut along $p_1$ through the node at $p_2=0$ and $p_3=\pi/2$. Fig.~\ref{fig:wh_system} (a) shows that the two states close to Weyl node momentum $p_1=0$ have group velocities of the same sign. The Brillouin zone-periodicity of bands thus requires additional low-energy states with opposite group velocities to appear at larger momenta. Those additional states must be taken into account when considering transport in a Weyl semimetal black or white hole analogue. Doing so is a fundamental difference of our microscopic lattice model to earlier low-momentum continuum models \cite{Volovik2016,Guan2017,Volovik2017,curved_spacetimes_ojanen_17,Huang2018,Zubkov2018,liu2019fermionic,Liang2019,Kedem2020,Hashimoto2020,stalhammar2021artificial}. As we discuss below, the additional doublers are in fact crucial for Hawking-like signatures Weyl systems.

\begin{figure}[t]
\captionsetup[subfloat]{labelformat=empty}
\captionsetup[subfloat]{farskip=0pt,captionskip=-100pt}
\subfloat[]{\includegraphics[width=0.9\columnwidth]{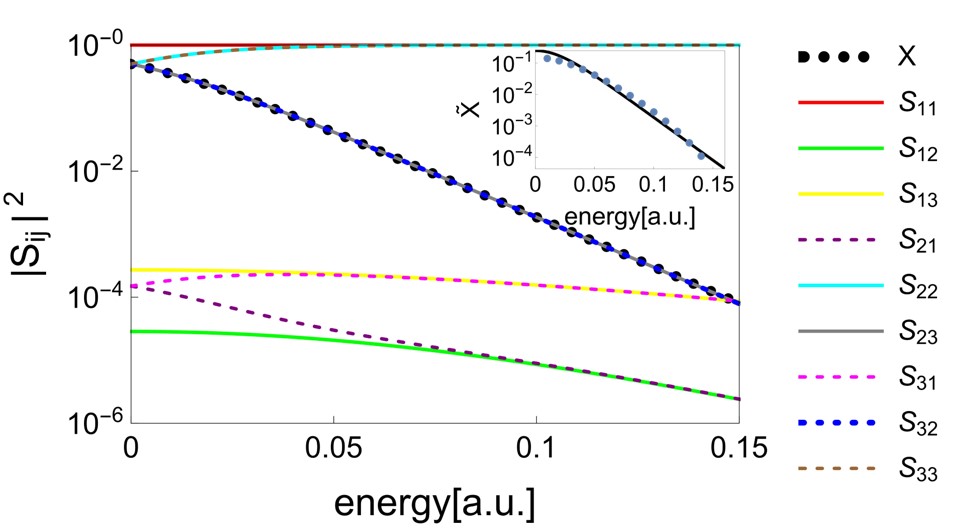}}
\vspace{-0.0in}\caption{Numerical results for our tight-binding white hole analogue at a cut along one node ($p_2=0$ and $p_3=\pi/a$), where we set $\alpha_{\rm t}=0.1$. As indicated by the dotted black line, the $S$-matrix elements $|S_{23}|^2$  and $|S_{32}|^2$ essentially equal the Hawking-like function $X$ of Eq.~\eqref{eq:hawking_expo}. Inset: the Hawking fragmentation rate proxy $\widetilde{X}$ of Eq.~\eqref{eq:Hawking_frac} (dots) closely matches $X\cdot(1-X)$ (solid line).\label{fig:BHTransProb}}
\end{figure}

Since cosmological black holes have smoothly varying frame fields (in Gullstrand-Painlev\'e coordinates), we focus on a Weyl semimetal black or white hole analogue with a smoothly position-dependent tilt $\vec{V}_{\rm t}$ (i.e.~small $\alpha_{\rm t}$ compared to the low-energy window; the sharp interface limit is physically very different and consequently has  distinct transport properties \cite{other_paper}). To obtain a microscopic tight-binding model, we translate Eq.~\eqref{eq:3DTBModel} to a real-space tight-binding mode with a spatially varying tilt. Concretely, we use $V_{\text{t},1}(x_1)=-1-\tanh\left(\alpha_{\rm t}\,x_1\right)$ for a black hole analogue, and $V_{\text{t},1}(x_1)=1+\tanh\left(\alpha_{\rm t}\,x_1\right)$ for a  white hole analogue. Close to the analogue horizon at $x_1=0$, this means $V_{\text{t},1}(x_1)\approx \mp(1+ \alpha_{\rm t}\,x_1)$. Comparing with Eq.~\eqref{eq:PGMEtricVelocity}, we find that the rate $\alpha_{\rm t}$ at which the tilt changes is analogous to surface gravity of a cosmological black hole, and thus to the effective Hawking temperature. Put simply, a spatially varying Weyl node tilt modifies group velocities of states in the node as a function of position, which maps to gravity changing $V(r)$ with $r$.

\emph{White hole analogue: scattering along a cut through a Weyl node. }
We begin by analyzing a cut along $p_1$ through one of the Weyl nodes by setting $p_2=0$ and $p_3=\pi/2$. Fig.~\ref{fig:wh_system} (a) illustrates the transition from an untilted node on the left to an overtilted node on the right due to $V_{\text{t},1}(x_1)=1+\tanh\left(\alpha_{\rm t}\,x_1\right)$.  Asymptotically far to the left, there are two states at energy $\omega$ close to the Weyl node with opposite group velocities. On the far right, there are four states at $\omega$: two low-momentum states with negative group velocity, and two high-momentum fermion doubler states with positive group velocity. Focussing only on the \emph{low-momentum} states, this setup can be identified as a white hole analogue. 

To understand how the low-energy doublers affect the analogy, we numerically simulate an inhomogenous one-dimensional tight-binding model of a Weyl semimetal white hole analogue along the cut $p_2=0$ and $p_3=\pi/2$ with KWANT \cite{Groth2014}. We focus on the scattering matrix $S$ that describes how incoming plane waves defined asymptotically far away from $x_1=0$ are scattered into asymptotic outgoing plane waves by the analogue event horizon. Labelling states as in Fig.~\ref{fig:wh_system} (a), the scattering matrix relates asymptotic incoming wavefunctions $\Psi_{\text{in}}\left(\omega\right)= \sum_{\nu=1}^3a_{\nu}\left(\omega\right)\psi_{\text{in},\nu}\left(\omega\right)$ to asymptotic outgoing wavefunctions
$\Psi_{\text{out}}\left(\omega\right)=\sum_{\mu=1}^3b_{\mu}\left(\omega\right)\psi_{\text{out},\mu}\left(\omega\right)$
as $b_\mu=S_{\mu\nu}\,a_\nu$. We numerically find that the $S$-matrix is approximately given by
\begin{align}
S&\approx\begin{pmatrix}1&0&0\\0&\sqrt{1-X}&\sqrt{X}\\0&\sqrt{X}&-\sqrt{1-X}\end{pmatrix},\label{eq:s-matrix}\\
X&=\frac{1}{1+e^{2\pi\omega/\alpha_{\rm t}}}\approx e^{-2\pi\omega/\alpha_{\rm t}}.\label{eq:hawking_expo}
\end{align}
Physically, this form of the $S$-matrix encodes three important pieces of information. First, for the band shown by a nearly straight dashed red line in Fig.~\ref{fig:wh_system} (a), the spatially varying tilt merely leads to a weakly renormalized group velocity. The central scattering region in which the node changes from untitled to overtilted is consequently transparent for states in this band: their transmission is $T\approx 1$. Second, the band shown by a solid blue curve in Fig.~\ref{fig:wh_system} (a) is strongly affected by the horizon. Its shape changes from approximately linear to S-like, and its number of states at low energies $\omega$ increases from one on the left to three on the right. The analogue event horizon induces strong scattering for states within this band, mixing the doubler states on the right lead with the non-doubler analogue states of the two leads, but we find virtually no scattering to states in the other band. Third and most importantly, the scattering is exponentially sensitive to how fast the tilt changes in the vicinity of the event horizon. Comparison with Eq.~\eqref{eq:hawking_probba} shows that the scattering probability $X$ takes a form analogous to the emission probability of Hawking radiation, with a Fermi function appearing in Eq.~\eqref{eq:hawking_expo} instead of the Bose function usual in bosonic black hole analogues. The unitarity of the $S$-matrix finally constrains the remaining probabilities to equal $1-X$. As shown in Fig.~\ref{fig:BHTransProb}, our microscopic lattice model only shows small deviations from the approximate form of Eq.~\eqref{eq:s-matrix}. Scattering between the red dashed band and the blue band is for example not strictly forbidden in our lattice model, but only occurs with negligible probability. To confirm that the observed scaling is universal, we  checked other tilt profiles with an approximately linear change at the horizon, and found the same exponential scaling of S-matrix elements.

\emph{Hawking fragmentation of wavepackets injected from a normal region into a white hole analogue. }
Inhomgenously tilted Weyl semimetals cannot emit Hawking-like radiation in equilibrium since they would otherwise have to evaporate like a black hole. The linear response regime close to equilibrium also does not show Hawking-like exponents: we numerically find the conductance to be $G\approx 1\,e^2/h$ \footnote{This statement is model dependent - other models realizing somewhat different forms of analogue event horizons can exhibit a non-trivial conductance. This will be detailed in a forthcoming publication.}. A suitable experiment probing the Hawking-analogue scattering probabilities $X$ of Eq.~\eqref{eq:hawking_expo} should provide access to particular scattering events. Hence, it requires preparing a Gaussian wavepacket $|\Phi_{\text{in}}\rangle=|\Phi_{\text{left}}^{(a_2)}\rangle$ centered around the state $a_2$ in the normal region, see Fig.~\ref{fig:wh_system} (a). Passing through the horizon, the wavepacket is fragmented into two outgoing wavepackets near the two fermion doubler states with probabilites $1-X$ and $X$. This is illustrated in Fig.~\ref{fig:wh_system} (b), and can be summarized as $|\Phi_{\text{in}}\rangle\to |\Phi_{\text{out}}\rangle=\sqrt{1-X}\, |\Phi_{\text{right}}^{(b_2)}\rangle+\sqrt{X}\, |\Phi_{\text{right}}^{(b_3)}\rangle$.

Since the system is spin-orbit-coupled, and because the outgoing fragments live at different momenta, they also have different spin polarizations. As shown in the Supplemental Material \cite{supplement}, this can be used to extract the Hawking fragmentation rate $X$ despite the fact that the outgoing wavepackets generally overlap in real space. We find that a suitable observable can be constructed from the spin expectation values $W_i = \langle \Phi_{\text{out}}|\sigma^i|\Phi_{\text{out}}\rangle$ and the total wavepacket weight $W_{\rm tot} = \langle \Phi_{\text{out}}|\Phi_{\text{out}}\rangle$ as

\begin{align}
\widetilde{X}= \frac{1}{2}\,\left(1-\frac{\sum_{i=1}^3W_i^2}{W_{\rm tot}^2}\right)\approx X\cdot(1-X)\cdot \xi.\label{eq:Hawking_frac}
\end{align}
Here, the outgoing wavepackets' spin-disalignement is encoded in $\xi$, a number of order unity, while the prefactor $X\cdot(1-X)$ measures wavepacket fragmentation. The inset of Fig.~\ref{fig:BHTransProb} shows the numerical evaluation of Eq.~\eqref{eq:Hawking_frac} in our one-dimensional lattice model. Our simulation confirm that $\widetilde{X}\approx X\cdot(1-X)$, thereby establishing $\widetilde{X}$ as a good marker for Hawking fragmentation. Alternatively, a spin filter or a spin-sensitive beam splitter can be used to separate the two transmission channels.

\emph{Hawking attenuation of wavepackets injected into a normal region from a black hole analogue. }
The $S$-matrix of a Weyl semimetal black hole analogue is identical to Eq.~\eqref{eq:s-matrix} modulo a relabelling of the states \cite{supplement}. Non-trivial scattering arises for a wavepacket prepared in the black hole region with states near one of the two large-momentum states (the fermion doublers). Depending on which of the two is chosen, the wavepacket is transmitted across the horizon with an amplitude reduced by a factor of $X$ or $1-X$, thereby giving rise to Hawking attenuation of the injected wavepacket \cite{supplement}.

\emph{Implications for three-dimensions and realizations. }
So far, we demonstrated our results on a one-dimensional cut through a Weyl node. This cut can be realized in metamaterials, which also provide exquisite control of the tilt profile. Weyl semimetal metamaterial analogues with tunable tilt have been reported in acoustic crystals \cite{Yang2016,Peri2019,Xie2019,Nejad2020}, photonics \cite{Ozawa2019}, topoelectrical circuits \cite{Rafi2020}, and cold atoms \cite{Xu2016}. Metamaterials also allow to measure spin and momentum resolved scattering and transport, and to study transport in specific transmission channels \cite{Lu2015,Cheng2020,Rafi2020}.


In a three-dimensional setup, Hawking fragmentation and Hawking attenuation can be probed similar to what is discussed above for the one-dimensional cut through the node if the wavepackets are very narrowly localized close to the node. In solids, the preparation of a controlled tilt profile is a key challenge. Possible solutions are carefully designed strain patterns \cite{Soluyanov2015,Guan2017,Alisultanov2019,MENG2020,Li2021,Nikolaev2021,Ferreira2021,Xie2021,jin2021}, and multilayer systems with specific inter-layer tunnelings \cite{Burkov2011,Sabsovich_2020}. In addition, the presence of additional states in three-dimensional solid or metamaterial analogues will generically impede the detection of Hawking fragmentation and Hawking attentuation. To more clearly isolate these effects, we propose to apply a magnetic field or pseudo-field  \cite{Vozmediano2015,Grushin2016,Pikulin2016,Ilan2019} along the direction of the tilt. These fields confine the low-energy states to (pseudo-)Landau levels that disperse only along the direction of the field. Crucially, the zeroth Landau level essentially behaves like the cut discussed above. (Pseudo-) magnetic fields in this sense help to highlight Hawking analogue physics, despite the fact that they are not directly related to them. To be most effective, the (pseudo-)magnetic fields should best be large: at small fields, the presence of higher Landau levels tends to mask the Hawking channels.

\begin{figure}[t]
\captionsetup[subfloat]{labelformat=empty}
\captionsetup[subfloat]{farskip=0pt,captionskip=-100pt}
\subfloat[]{\includegraphics[width=0.94\columnwidth]{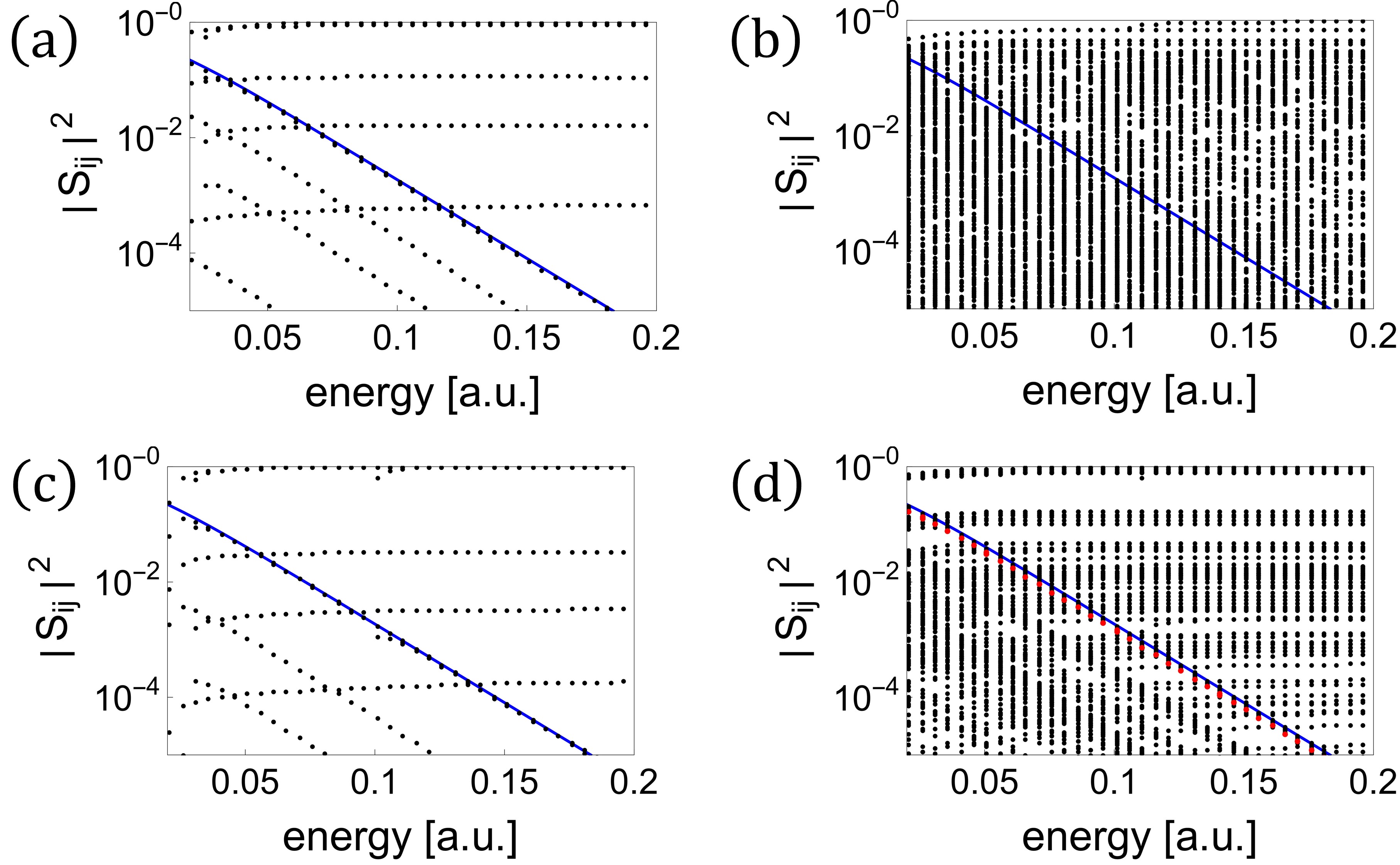}}
\vspace{-0.0in}\caption{Transmission elements $|S_{ij}|^2$ of the $S$-matrix for three-dimensional Weyl semimetal black hole analogues in nanowires, where we set $\alpha_{\rm t}=0.1$. Panels (a) and (b) are without, (c) and (d) with magnetic pseudofield for crossections of $5\times5$ [(a), (c)] and $15\times15$ [(b), (d)] sites. Solid blue lines show $X$ from Eq.~\eqref{eq:hawking_expo}. In (d) the red line is an example transmission element following $X$. \label{fig:3DProb}\vspace{-0.2in}}
\end{figure}

Since Weyl node tilt profiles can be created by inhomogenous strain, and given that strain can also give rise to magnetic pseudofields, we here focus on the latter. Fig.~\ref{fig:3DProb} shows the transmission elements of the $S$-matrix calculated numerically for our tight-binding model of a Weyl semimetal black hole analogue in a nanowire geometry with open boundary conditions. For thin wires ($5\times5$ sites), the Hawking fragmentation rate is clearly visible both without (panel (a)) and with (panel (c)) a pseudofield. In thicker nanowires, the Hawking fragmentation rate is hidden by the strongly increased number of scattering channels without pseudofield (panel (c)). The application of a magnetic pseudofield resurfaces the Hawking fragmentation rate (panel (d)).

\emph{Conclusions. } In this work, we analyzed analogues of black and white holes in Weyl semimetals. The analogy derives from a mapping of the Weyl Hamiltonian, an approximation to the full Hamiltonian at small momenta, to the Schwarzschild metric in Gullstrand-Painlev\'e coordinates. We find that the additionally low-energy states appearing in the lattice system are key for the observation of the analogue of a Hawking temperature. The scattering of wavepackets constructed from these additional low-energy states by the analogue event horizon depends exponentially on the analogue Hawking temperature, which microscopically is set by the rate at which the tilt changes in space. This exponential sensitivity of wave packet transport not only ties a new connection between solid state lattice models and the physics of black holes, but also open the door to applications of Weyl semimetal black and white hole analogues using this exponential sensitivity for, \emph{e.g.}, sensing.\newline


\begin{acknowledgments}
The authors thank Christian Schmidt and Piet Brouwer for invaluable discussions. RI and DS are supported by the Israel Science Foundation (ISF, Grant No.1790/18). PW and TM acknowledge financial support by the Deutsche Forschungsgemeinschaft via the Emmy Noether Programme ME4844/1-1 (project id 327807255), the Collaborative Research Center SFB 1143 (project id 247310070), and the Cluster of Excellence on Complexity and Topology in Quantum Matter ct.qmat (EXC 2147, project id 390858490).

\emph{Note:} A related publication \cite{other_paper} provides an analytical derivation of the $S$-matrix, and studies non-equilibrium effects in three-dimensional solid-state Weyl semimetal black hole analogues.
\end{acknowledgments}

\bibliography{draft}

\appendix

\section{A few simplistic  estimates.}

\paragraph{Speed of a light ray in Gullstrand-Painlev\'e-coordinates.}
As discussed in the main text, the Schwarzschild metric in Gullstrand-Painlev\'e-coordinates features a velocity 
\begin{align}
\vec{V}(r)&={-c\sqrt{\frac{r_{h}}{r}}}\,\vec{e}_r=-V(r)\,\vec{e}_r
\end{align}
that describes the speed of a free-falling object. This velocity diverges for $r\to0$, and equals $\vec{V}(r) = -c\,\vec{e}_r$ at the horizon. This does not mean that the object becomes superluminal within the horizon, but is merely a curiosity of the chosen coordinates. To appreciate this point, recall that the propagation of a light ray is defined by $ds^2=0$. From the Schwarzschild metric in Gullstrand-Painlev\'e-coordinates, this implies

\begin{align}
c^2 &= \left(\frac{d\vec{r}}{dt}-\vec{V}\right)^2.
\end{align}
Since the motion is radial, we have
\begin{align}
\frac{d\vec{r}}{dt}=\vec{v}_{\text{lightray}}= v_{\text{lightray}}\,\vec{e}_r,
\end{align}
and thus $v_{\text{lightray}} = \pm c -V(r)$. The resulting velocities
\begin{align}
\vec{v}_{\text{lightray},\pm} =  \pm c\,\vec{e}_r+ \vec{V}(r)\label{eq:append_lightray}
\end{align}
correspond to the velocities of a light ray emitted from a torch at a distance $r$ from the origin (the black hole resides at $r=0$) pointing inwards or outwards. For $r\to\infty$, where $V(r)\to0$, the velocities approach $\pm c\,\vec{e}_r$. At the horizon, the velocity $\vec{v}_{\text{lightray},+}$ of a lightray emitted outwards vanishes. For $r<r_{\text{hor}}$, finally, both velocities point towards the center of the black hole. 

The velocities $\vec{v}_{\text{lightray},\pm} $ define a cut through the lightcone at the point of the torch in the radial direction. Clearly, a free-falling object moving with the velocity $V(r)\,\vec{e}_r$ always stays within that lightcone, is is thus slower than the speed of light in Gullstrand-Painlev\'e-coordinates, as it should be.

\paragraph{A simple classical picture for accelerated motion.} Motion with a constant acceleration is described by the Rindler coordinates, which take into account the fact that velocity cannot exceed the speed of light. However, at small enough acceleration and time-span we can apply a simpler procedure. Consider an object that at time $t=0$ resides at the position $\vec{r}_0 = r_0\,\vec{e}_r$ and has the initial radial velocity $\vec{v}_0 = v_0\,\vec{e}_r$. For $t>0$, the object is accelerated by a radial force $\vec{F} = m\,a\,\vec{e}_r$. Considering the motion for $t>0$ as Newtonian, we find
\begin{align}
\vec{r}(t)&=\left(r_0+v_0\,t+\frac{1}{2}\,a\,t^2\right)\,\vec{e}_r,\\
\vec{v}(t)&=\left(v_0+a\,t\right)\,\vec{e}_r.
\end{align}
We can now express time as a function of the $r$-coordinate. For small enough times such that $v_0\,t\gg a\,t^2$, we have
\begin{align}
\vec{r}(t)=r(t)\,\vec{e}_r&\approx\left(r_0+v_0\,t\right)\,\vec{e}_r.\\
\end{align}
This implies
\begin{align}
t\approx\,\frac{r-r_0}{v_0}~~\Rightarrow~~\vec{v}(t)=\left(v_0+\frac{a}{v_0}\,(r-r_0)\right)\,\vec{e}_r.\label{eq:appendmotion}
\end{align}
In the main text, the acceleration stems from gravity, and the role of $v_0$ at the horizon is played by $-V(r)=-c$. Comparison of \eqref{eq:appendmotion} with Eq.~(2) of the main text shows that the parameter $\alpha$ there is naturally related to the surface gravity of the black hole. For further comparison, recall that the surface gravity is $g_{\text{surf}}=G\,M_{\bullet}/r_{\text{hor}}^2={c^4}/{4\,G\,M_{\bullet}}=\alpha\,c$.

\paragraph{Tilted lightcones and tilted Weyl nodes.}
Consider the Hamiltonian describing a tilted Weyl node given in Eq.~(6) of the main text. The two eigenvalues of this Hamiltonian are
\begin{align}
E_{1}(\vec{p})&= v_F\,|\vec{p}|+\vec{V}_{\text{t}}\cdot\vec{p},\\
E_{2}(\vec{p})&= -v_F\,|\vec{p}|+\vec{V}_{\text{t}}\cdot\vec{p}.
\end{align}
The group velocities associated with these two bands are
\begin{align}
\vec{v}_{1}(\vec{p})&= v_F\,\frac{\vec{p}}{|\vec{p}|}+\vec{V}_{\text{t}}= v_F\,\vec{e}_p+\vec{V}_{\text{t}},\\
\vec{v}_{2}(\vec{p})&=  -v_F\,\frac{\vec{p}}{|\vec{p}|}+\vec{V}_{\text{t}}= -v_F\,\vec{e}_p+\vec{V}_{\text{t}}.
\end{align}
These velocities are analogous of the tilted lightcone velocities in Eq.~\eqref{eq:append_lightray} with the Fermi velocity replacing the speed of light and with $\vec{V}_{\text{t}}$ taking the place of $\vec{V}$. Let us at this point mention a subtlety arising in the connection between tilted Weyl nodes and tilted lightcones: the slope of lightcones is defined by the inverse of the velocities discussed above, since a lightcone can be understood as "time as a function of spatial coordinates'', and time equals position \emph{divided by} velocity. The slope of a Weyl cone, on the contrary, is defined by a velocity (energy equals momentum \emph{times} velocity). We thus see that the slopes of the lightcone along the radial direction (defined with respect to the black hole) equals the inverse of the slope of a tilted Weyl node.

\section{Black hole analogues and Hawking attenuation: additional details.}
\begin{figure}[t]
\captionsetup[subfloat]{labelformat=empty}
\captionsetup[subfloat]{farskip=0pt,captionskip=-100pt}
\raisebox{4.1cm}{(a)}\subfloat[]{\includegraphics[width=0.9\columnwidth]{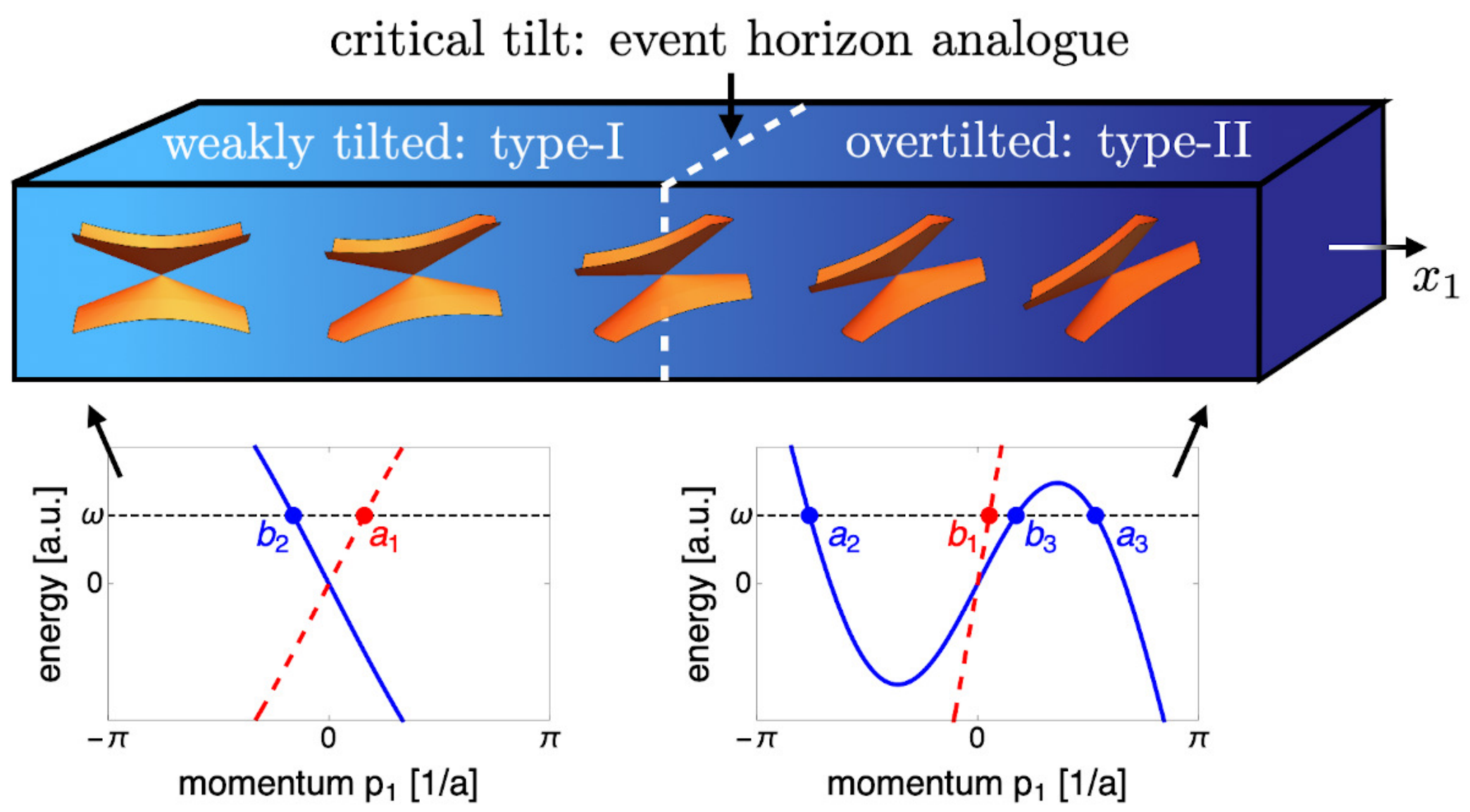}}\\[0.75cm]
\raisebox{2.15cm}{(b)}\subfloat[]{\includegraphics[width=0.85\columnwidth]{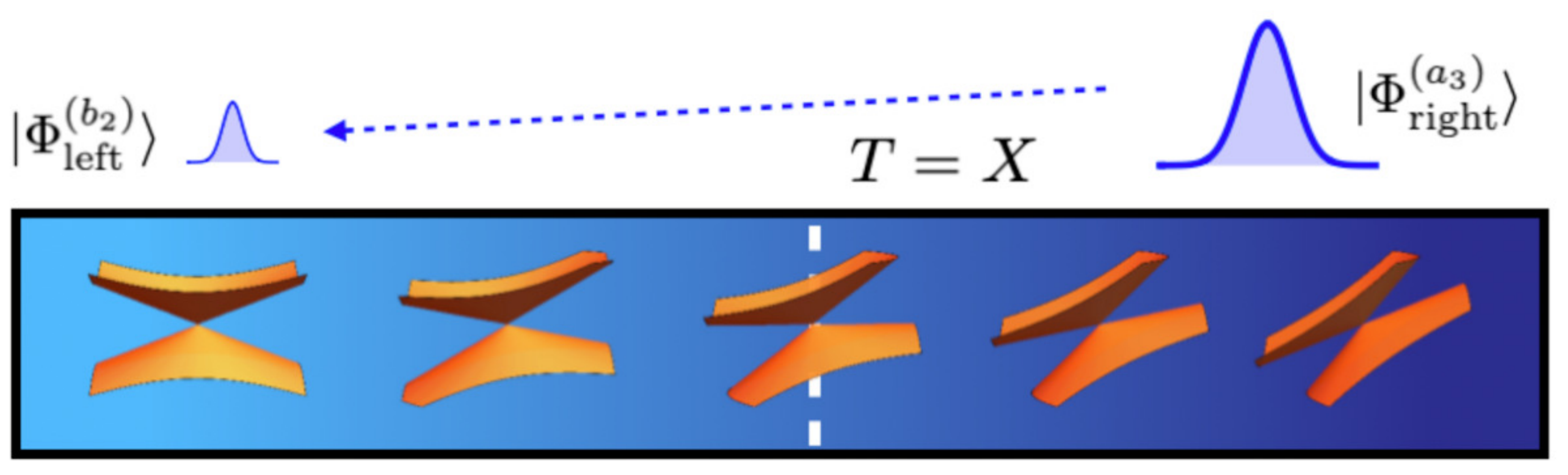}\quad}
\vspace{-0.0in}\caption{Panel (a): Black hole analogue realized in a Weyl semimetal with inhomogenous nodal tilt. Panel (b): sketch of Hawking attentuation. All lables are similar to Fig.~1 in the main text.\label{fig:bh_system}}
\end{figure}

A black hole analogue can be constructed by tilting the node towards the horizon. As shown in panel (a) of Fig.~\ref{fig:bh_system}, the low-momentum modes then correspond to states moving away from the analogue horizon. The modes in the dispersions defined by the asymptotic Hamiltonian have a modified labeling in order to bring the $S$-matrix to the same form as in Eq.~(8) of the main text. Panel (b) sketches Hawking attenuation: a wavepacket prepared around the state $a_3$ within the ``black hole region'' is transmitted to mode the ``normal'' side at a reduced amplitude. The remaining spectral weight is reflected into a wavepacket centered around the state $b_3$.

\section{Wavepacket scattering and the $S$-matrix}
To calculate wavepacket scattering for the one-dimensional cut through a white hole analogue discussed in the main text, we begin by preparing a Gaussian wavepacket centered around the state labeled by $a_2$ in Fig.~(1) (a) of the main text. The momentum of this state is $p_1=p_{a_2}$. To that end, we use the Gaussian  
\begin{eqnarray}
g_{p_{a_2}}(p_1)=\frac{1}{2\sqrt{\pi}s}e^{-\frac{(p_1-p_{a_2})^{2}}{4s^{2}}} ,\label{eq:GaussianDist}
\end{eqnarray}
where $s$ sets the width of the wavepacket (which we will choose to be narrow in momentum space). The wavepacket has the real space wavefunction
\begin{eqnarray}
\langle x_1|\Phi_{\text{left}}^{(a_2)}\rangle=\frac{1}{\sqrt{V}}\sum_{p_1}e^{ip_1x_1}\,g_{p_{a_2}}(p_1)\,\psi_{a_2}(p_1),
\end{eqnarray}
where $V$ is the system's volume in real space, and $\psi_{a_2}(p_1)$ is an asymptotic incoming spinor wavefunction at momentum $p_1$ in the same band as the state $a_2$, see Fig.~(1) (a) of the main text. Via the dispersion $\omega_{a_2}(p_1)$, each state in the band is associated with an energy $\omega$. This allows to identify 
\begin{align}
\psi_{a_2}(p_1)\equiv\psi_{\rm in,2}(\omega)\quad\text{and}\quad g(p_1-p_{a_2})\equiv a_2(\omega),\label{eq:coeff-decompose}
\end{align}
thereby rewriting the component of the incoming wavepacket at energy $\omega$ as in the main text as $ \Psi_{\text{in}}\left(\omega\right)= a_{2}\left(\omega\right)\psi_{\text{in},2}\left(\omega\right)$.

\begin{figure}[t]
\captionsetup[subfloat]{labelformat=empty}
\captionsetup[subfloat]{farskip=0pt,captionskip=-100pt}
\raisebox{2.cm}{(a)}\subfloat[]{\includegraphics[width=0.4\columnwidth]{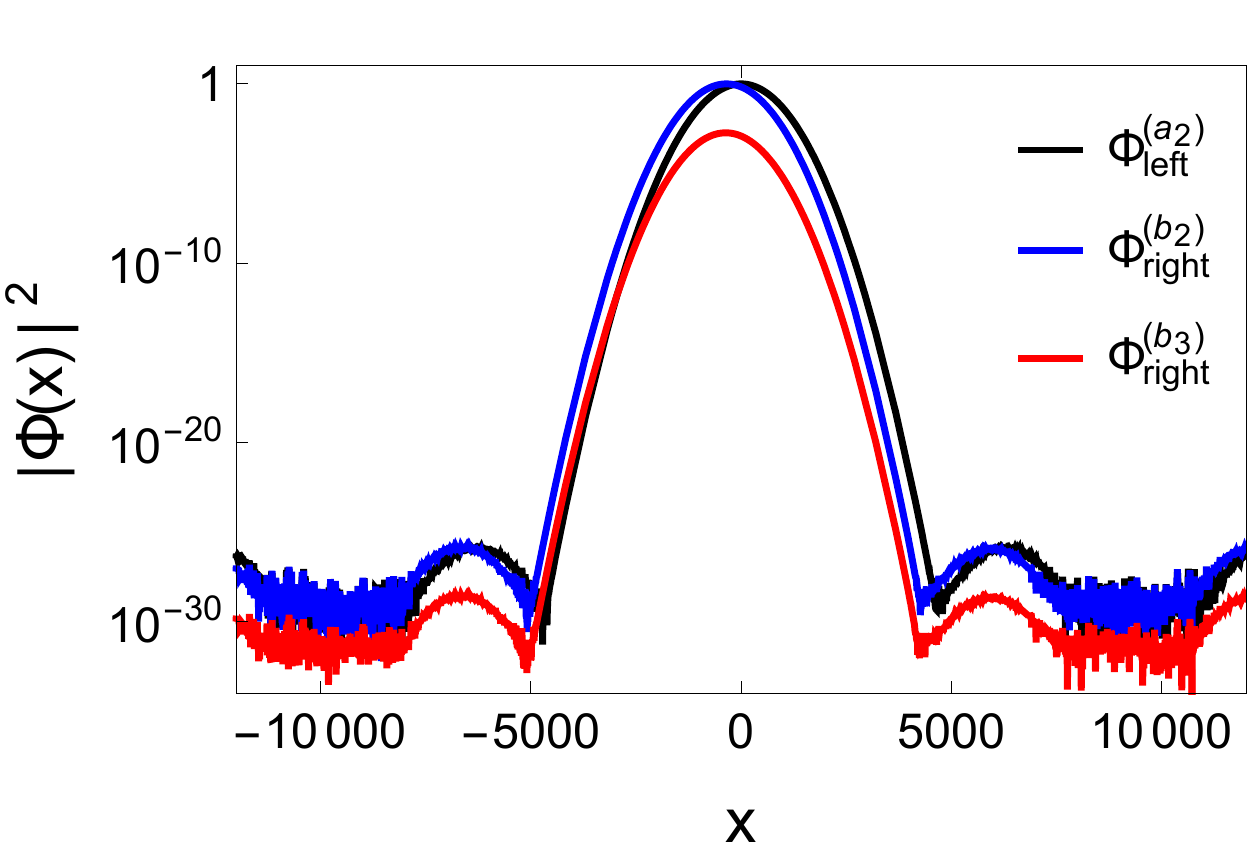}}\quad
\raisebox{2.cm}{(b)}\subfloat[]{\includegraphics[width=0.4\columnwidth]{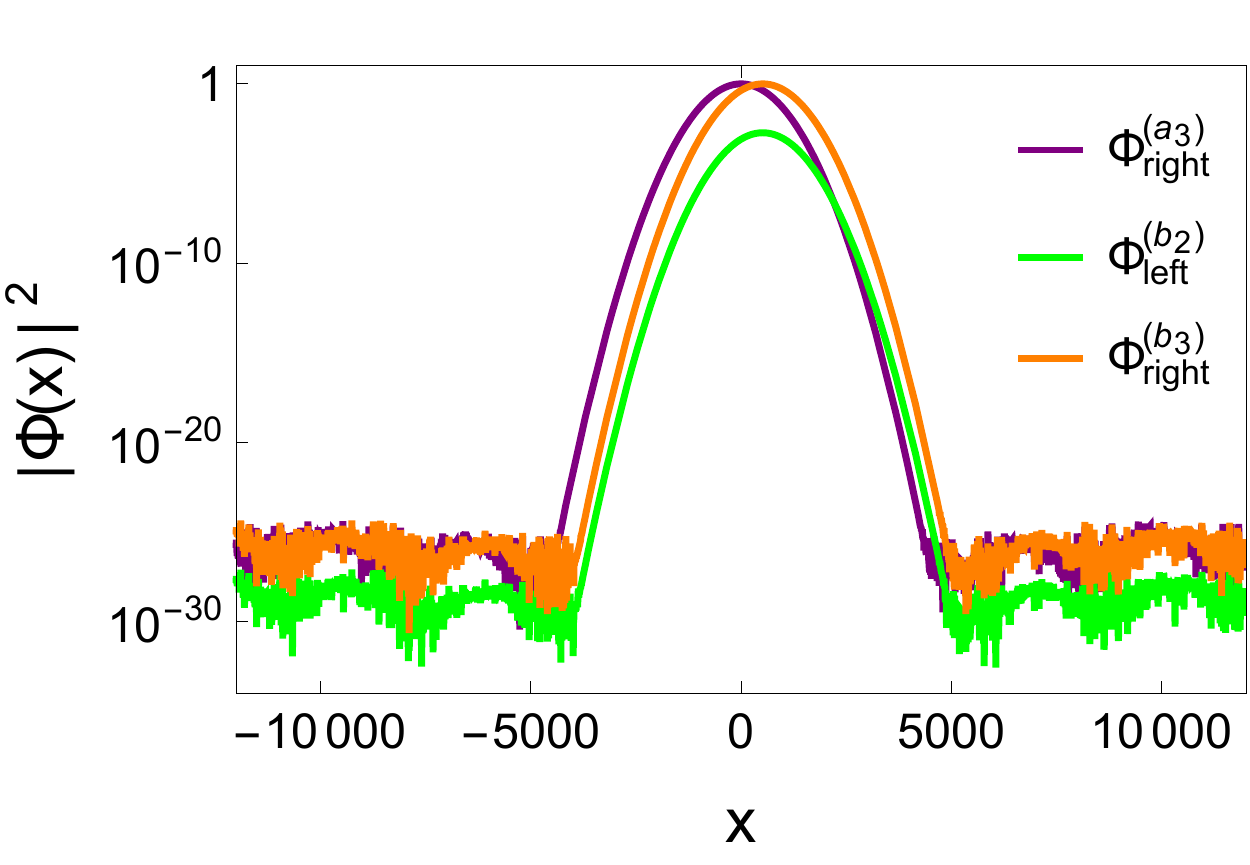}}
\vspace{-0.0in}\caption{Ingoing and outgoing wavepackets in real space after scattering through the white hole (a) or black hole (b) horizon. For (a) the incoming wavepacket is centered around $p_1=-0.1$ with a standard deviation $s=0.00125$. For (b) the incoming wavepacket is cantered around $p_1=2.026$ with a standard deviation $s=0.00125$. For both the white and black hole horizons the length of the scattering region is $L=400$, and we use $\alpha_{\rm t}=0.1$. The tilt amplitude changes between $V_{t,1}=0$ at the left end of the scattering region, and $V_{t,1}=\pm2$ at the right end of the scattering region. \label{fig:WavePackets}}
\end{figure}

To compute the outgoing wavepackets, we utilize the $S$-matrix, which relates amplitudes of incoming plane waves defined asymptotically far away from the horizon  to  amplitudes of the asymptotic outgoing plane waves. The $S$-matrix can be understood as a time evolution operator in the interaction picture from $t=-\infty$, where we define our incoming state $\left|\Phi_{\rm in}\right\rangle$, to $t=+\infty$, where we want to measure the outgoing states $\left|\Phi_{\rm out}\right\rangle$ (with the central scattering region defining the ``perturbation Hamiltonian").
For the white hole analogue depicted in Fig.~(1) (b) of the main text, the incoming wavepacket fragments into two right-moving wavepackets centered around the states $b_2$ and $b_3$ in the type-II region. The corresponding amplitudes at energy $\omega$ are

\begin{eqnarray}
b_2(\omega) = S_{22}(\omega)\,a_{2}(\omega)~~\text{and}~~ b_3(\omega) = S_{32}(\omega)\,a_{2}(\omega),
\end{eqnarray}
from which we construct the outgoing wavepacket by multiplying the coefficients $b_{2,3}$ with the relevant outgoing wavefunctions (plane wave times spinor), and then summing the components for different energies $\omega$. We plot the incoming and outgoing wavepackets in Fig. \ref{fig:WavePackets} (a), which shows that both outgoing wavepacket fragments have approximately the same shape as the incoming wavepacket.

The wavepacket scattering through the black hole horizon follows the same steps. The incoming wavepacket is constructed with a Gaussian distribution around the state labelled by $a_3$ in Fig.~\ref{fig:bh_system} (a). Using the $S$-matrix, we then calculate the scattering described in Fig.~\ref{fig:bh_system} (b) of the incoming wavepacket into outgoing wavepackets, see Fig.~\ref{fig:bh_system} (a). We plot the incoming and outgoing wavepackets in Fig.~\ref{fig:WavePackets} (b), 

\section{The Hawking fragmentation rate proxy $\tilde{X}$}
The Hawking fragmentation rate proxy $\widetilde{X}$ discussed in the main text is a rather generic quantity fingerprinting superimposed wavepackets in spin-orbit coupled systems. In particular, it is not specific to Weyl systems (but Hawking fragmentation is). To motivate $\widetilde{X}$, we consider a general two-band Bloch Hamiltonian $\mathcal{H}_{\vec{p}}=\mathcal{H}_{\vec{p}}^\dagger$, which can be decomposed into Pauli matrices as
\begin{align}
\mathcal{H}_{\vec{p}} = h_{0,\vec{p}}\,\mathds{1}+\sum_{i=1}^3 h_{i,\vec{p}}\,\sigma^i\quad\text{with}\quad h_{0,\vec{p}},h_{i,\vec{p}}\in\mathds{R}.
\end{align}
Using spherical coordinates for the momenta, $\mathcal{H}_{\vec{p}}$ can be rewritten as

\begin{align}
\mathcal{H}_{\vec{p}} = h_{0,\vec{p}}\,\mathds{1}+h_{\vec{p}}\,\bigl(&\cos(\phi_{\vec{p}})\sin(\theta_{\vec{p}})\,\sigma^1+\sin(\phi_{\vec{p}})\sin(\theta_{\vec{p}})\,\sigma^2\nonumber\\
&+\cos(\theta_{\vec{p}})\,\sigma^3\bigr)
\end{align}
with $h_{\vec{p}} = \sqrt{\sum_{i=1}^3 h_{i,\vec{p}}^2}$. The orthonormalized eigenvectors $\vec{\psi}_{\vec{p},\pm}$ of $\mathcal{H}_{\vec{p}}$ are

\begin{align}
\vec{\psi}_{\vec{p},+}=\begin{pmatrix}e^{-i\phi_{\vec{p}}}\,\cos(\theta_{\vec{p}}/2)\\\sin(\theta_{\vec{p}}/2)\end{pmatrix},\\
\vec{\psi}_{\vec{p},-}=\begin{pmatrix}-e^{-i\phi_{\vec{p}}}\,\sin(\theta_{\vec{p}}/2)\\\cos(\theta_{\vec{p}}/2)\end{pmatrix}.
\end{align}
The associated eigenvalues are $E_{\vec{p},\pm}=h_{0,\vec{p}}\pm h_{\vec{p}}$. The spin-expectation values of these eigenstates are
\begin{align}
\vec{\psi}_{\vec{p},\pm}^\dagger\,\sigma^1\,\vec{\psi}_{\vec{p},+}^\pdag &= \pm\cos(\phi_{\vec{p}})\sin(\theta_{\vec{p}}),\\
\vec{\psi}_{\vec{p},\pm}^\dagger\,\sigma^2\,\vec{\psi}_{\vec{p},+}^\pdag &= \pm\sin(\phi_{\vec{p}})\sin(\theta_{\vec{p}}),\\
\vec{\psi}_{\vec{p},\pm}^\dagger\,\sigma^3\,\vec{\psi}_{\vec{p},+}^\pdag &= \pm\cos(\theta_{\vec{p}}).
\end{align}
Now consider an initial wavepacket $|\Phi_{0,\pm}\rangle$ defined by

\begin{align}
\langle\vec{r}|\Phi_{0,\pm}\rangle=\Phi_{0,\pm}(\vec{r})=\frac{1}{\sqrt{V}}\sum_{\vec{p}}e^{i\vec{p}\cdot\vec{r}}\,f_{\vec{p}_0}(\vec{p})\,\vec{\psi}_{\vec{p},\pm},
\end{align}
where $V$ is the system's real-space volume. The envelope function $f_{\vec{p}_0}(\vec{p})$ is centered at $\vec{p}_0$, and may for example be a Gaussian. The total weight in this initial wavepacket is
\begin{align}
&W_{\rm tot}=\langle\Phi_{0,\pm}|\Phi_{0,\pm}\rangle \nonumber\\
&= \int d^3r\,\frac{1}{{V}}\sum_{\vec{p},\vec{q}}e^{-i(\vec{p}-\vec{q})\cdot\vec{r}}\,f_{\vec{p}_0}^*(\vec{p})\,f_{\vec{p}_0}^\pdag(\vec{q}) \vec{\psi}_{\vec{p},\pm}^\dagger\,\vec{\psi}_{\vec{q},\pm}.
\end{align}
Using the representation of the Kronecker delta in terms of exponentials, and the orthonormalization of the eigenstates, this yields 
\begin{align}
W_{\rm tot}&= \sum_{\vec{p},\vec{q}}\,{\delta}_{\vec{p},\vec{q}}\,f_{\vec{p}_0}^*(\vec{p})\,f_{\vec{p}_0}^\pdag(\vec{q}) \,\vec{\psi}_{\vec{p},\pm}^\dagger\,\vec{\psi}_{\vec{q},\pm}\nonumber\\
&= \sum_{\vec{p}}|f_{\vec{p}_0}^\dagger(\vec{p})|^2.
\end{align}
This wavepacket is now assumed to be scattered into two fragments $|\Phi_{1,\pm}\rangle$ centered around $\vec{p}_1$ and $|\Phi_{2,\pm}\rangle$ centered around $\vec{p}_2$ with probabilities $T_1$ and $T_2=1-T_1$, respectively. We furthermore assume that the shape of the two final wavepackets is identical to the initial wavepacket's form. For the state, this translates to $|\Phi_{0,\pm}\rangle\to|\Phi_{\text{final},\tau\tau'}\rangle$ with
\begin{align}
\Phi_{\text{final},\tau\tau'}(\vec{r}) = &\frac{\sqrt{T_1}}{\sqrt{V}}\sum_{\vec{p}}e^{i\varphi_{1,\vec{p}}}\,e^{i\vec{p}\cdot\vec{r}}\,f_{\vec{p}_1}(\vec{p})\,\vec{\psi}_{\vec{p},\tau}\nonumber\\
&+\frac{\sqrt{1-T_1}}{\sqrt{V}}\sum_{\vec{p}}e^{i\varphi_{2,\vec{p}}}\,e^{i\vec{p}\cdot\vec{r}}\,f_{\vec{p}_2}(\vec{p})\,\vec{\psi}_{\vec{p},\tau'},
\end{align}
where $\varphi_{1,2}$ are phase factors arising from the scattering, and $\tau=\pm$ is independent from $\tau'=\pm$. Furthermore, we assume that the two final wavepacket-parts do not overlap. That means that $f_{\vec{q}}(\vec{p})$ is narrowly centered around $\vec{q}$, and that $\vec{p}_1-\vec{p}_2$ is much larger than the width of $f_{\vec{q}}(\vec{p})$. For an observable $\mathcal{O}$, the expectation value with respect to $|\Phi_{\text{final},\pm}\rangle$ is
\begin{widetext}
\begin{align}
&\langle\Phi_{\text{final},\tau\tau'}|\mathcal{O}|\Phi_{\text{final},\tau\tau'}\rangle=\int d^3r\,\Phi_{\text{final},\tau\tau'}^\dagger(\vec{r})\,\mathcal{O}\,\Phi_{\text{final},\tau\tau'}^\pdag(\vec{r})\nonumber\\
&=\sum_{\vec{p}}\,\Biggl(T_1\,f_{\vec{p}_1}^*(\vec{p})\,f_{\vec{p}_1}^\pdag(\vec{p})\,\vec{\psi}_{\vec{p},\tau}^\dagger\,\mathcal{O}\,\vec{\psi}_{\vec{p},\tau} +(1-T_1)\,f_{\vec{p}_2}^*(\vec{p})\,f_{\vec{p}_2}^\pdag(\vec{p})\,\vec{\psi}_{\vec{p},\tau'}^\dagger\,\mathcal{O}\,\vec{\psi}_{\vec{p},\tau'}\nonumber\\
&\hspace{1.5cm}+\sqrt{T_1(1-T_1)}\,(e^{-i(\varphi_{1,\vec{p}}-\varphi_{2,\vec{p}})}\, \underbrace{f_{\vec{p}_1}^*(\vec{p})\,f_{\vec{p}_2}^\pdag(\vec{p})}_{\approx0~\text{(no overlap)}}\,\vec{\psi}_{\vec{p},\tau}^\dagger\,\mathcal{O}\,\vec{\psi}_{\vec{p},\tau'}+\text{h.c.})\Biggr)\nonumber\\
&\approx\sum_{\vec{p}}\,T_1\,|f_{\vec{p}_1}(\vec{p})|^2\,\vec{\psi}_{\vec{p},\tau}^\dagger\,\mathcal{O}\,\vec{\psi}_{\vec{p},\tau}+\sum_{\vec{p}}\,(1-T_1)\,|f_{\vec{p}_2}(\vec{p})|^2\,\vec{\psi}_{\vec{p},\tau'}^\dagger\,\mathcal{O}\,\vec{\psi}_{\vec{p},\tau'}.
\end{align}
As a final assumption, we consider the wavepackets' width, set by $f_{\vec{p}_i}(\vec{p})$, to be much narrower than the momentum scale along which the Hamiltonian, and thus the eigenstates, change. Within each wavepacket, the  spinors $\vec{\psi}_{\vec{p},\nu}$ can thus be approximated as constants. This yields

\begin{align}
\langle\Phi_{\text{final},\tau\tau'}|\mathcal{O}|\Phi_{\text{final},\tau\tau'}\rangle&\approx\vec{\psi}_{\vec{p}_1,\tau}^\dagger\,\mathcal{O}\,\vec{\psi}_{\vec{p}_1,\tau}\,\sum_{\vec{p}}\,T_1\,|f_{\vec{p}_1}(\vec{p})|^2+\vec{\psi}_{\vec{p}_2,\tau'}^\dagger\,\mathcal{O}\,\vec{\psi}_{\vec{p}_2,\tau'}\,\sum_{\vec{p}}\,(1-T_1)\,|f_{\vec{p}_2}(\vec{p})|^2\nonumber\\
&=W_{\rm tot}\,\left(T_1\,\vec{\psi}_{\vec{p}_1,\tau}^\dagger\,\mathcal{O}\,\vec{\psi}_{\vec{p}_1,\tau}+(1-T_1)\,\vec{\psi}_{\vec{p}_2,\tau'}^\dagger\,\mathcal{O}\,\vec{\psi}_{\vec{p}_2,\tau'}\right).
\end{align}
The total weight of the final wavepacket is obtained by setting $\mathcal{O}\to\mathds{1}$, and yields
\begin{align}
\langle\Phi_{\text{final},\tau\tau'}|\mathds{1}|\Phi_{\text{final},\tau\tau'}\rangle = W_{\rm tot}\,\left(T_1+(1-T_1)\right)=W_{\rm tot}.
\end{align}
The total weight is thus conserved in the scattering, as it should be. Using the abbreviations
\begin{align}
x_i = \cos(\phi_{\vec{p}_i})\sin(\theta_{\vec{p}_i})\quad\text{and}\quad y_i= \sin(\phi_{\vec{p}_i})\sin(\theta_{\vec{p}_i})\quad\text{and}\quad z_i= \cos(\theta_{\vec{p}_i}),
\end{align}
the spin expectation values yield
\begin{align}
\langle\Phi_{\text{final},\tau\tau'}|\sigma^1|\Phi_{\text{final},\tau\tau'}\rangle^2 &= W_{\rm tot}^2\,\left(T_1^2\,x_1^2+(1-T_1)^2\,x_2^2+2\,\tau\,\tau'\,T_1\,(1-T_1)\,x_1\,x_2\right),\\
\langle\Phi_{\text{final},\tau\tau'}|\sigma^2|\Phi_{\text{final},\tau\tau'}\rangle^2 &= W_{\rm tot}^2\,\left(T_1^2\,y_1^2+(1-T_1)^2\,y_2^2+2\,\tau\,\tau'\,T_1\,(1-T_1)\,y_1\,y_2\right),\\
\langle\Phi_{\text{final},\tau\tau'}|\sigma^3|\Phi_{\text{final},\tau\tau'}\rangle^2 &= W_{\rm tot}^2\,\left(T_1^2\,z_1^2+(1-T_1)^2\,z_2^2+2\,\tau\,\tau'\,T_1\,(1-T_1)\,z_1\,z_2\right),
\end{align}
where we recall that $\tau,\tau'=\pm\equiv\pm1$. Using $x_i^2+y_i^2+z_i^2=1$, we find
\begin{align}
\sum_{i=1}^3\langle\Phi_{\text{final},\tau\tau'}|\sigma^i|\Phi_{\text{final},\tau\tau'}\rangle^2\approx W_{\rm tot}^2\,\left(1-2\,T_1(1-T_1)+2\,T_1\,(1-T_1)\,\tau\,\tau'\,(x_1\,x_2+y_1\,y_2+z_1\,z_2)\right)
\end{align}
Defining
\begin{align}
{\xi}=1-\tau\,\tau'\,(x_1\,x_2+y_1\,y_2+z_1\,z_2),
\end{align}
we finally obtain
\begin{align}
1-\frac{\sum_{i=1}^3\langle\Phi_{\text{final},\tau\tau'}|\sigma^i|\Phi_{\text{final},\tau\tau'}\rangle^2}{W_{\rm tot}^2}&\approx 2\,T_1(1-T_1)\,{\xi}.
\end{align}
Eq.~(10) from the main text then follows by identifying $T_1=X$.
\end{widetext}

\section{Other tilt profiles}
\begin{figure}[t]
\captionsetup[subfloat]{labelformat=empty}
\captionsetup[subfloat]{farskip=0pt,captionskip=-100pt}
\subfloat[]{\includegraphics[width=0.9\columnwidth]{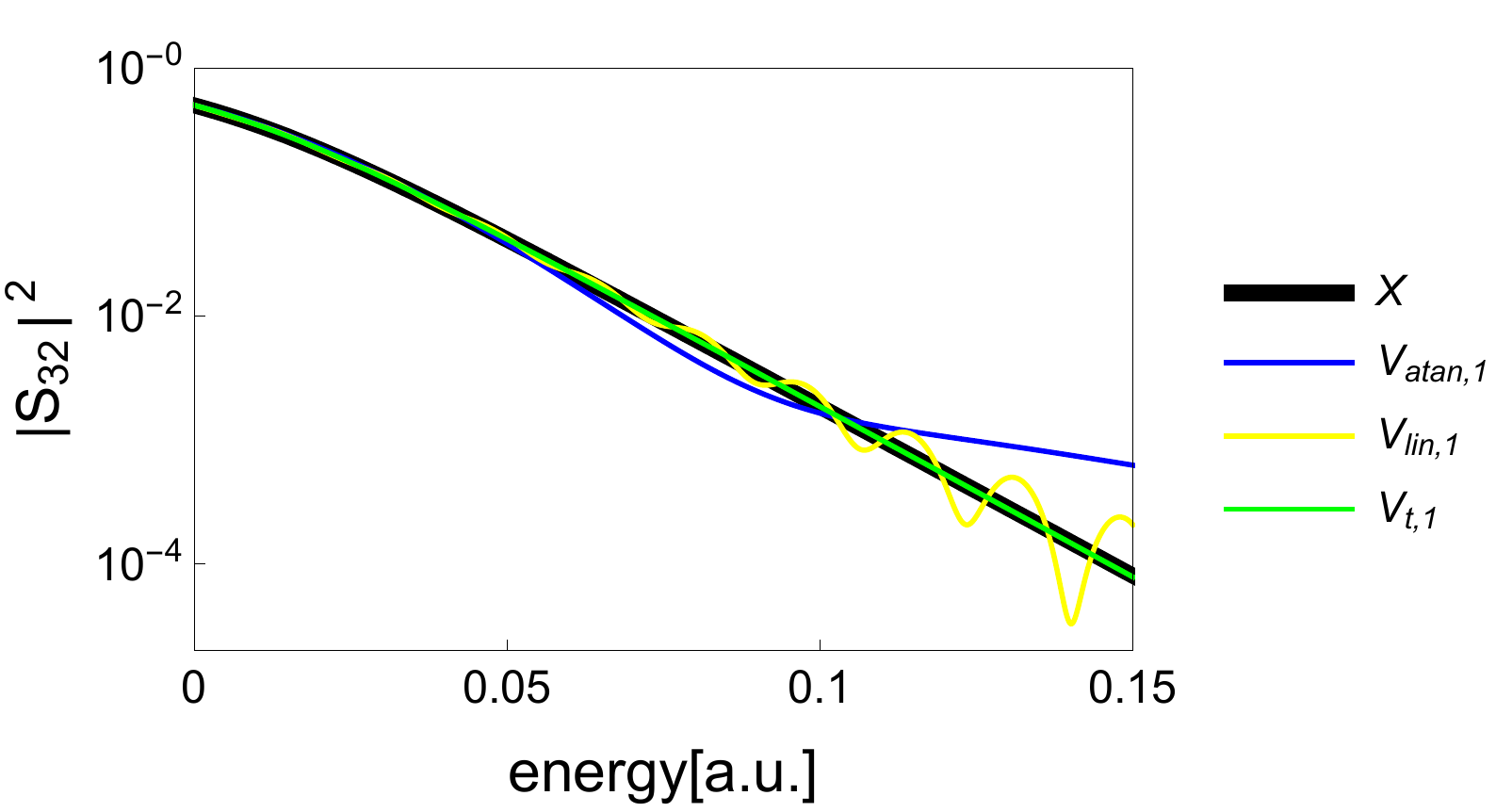}}
\vspace{-0.0in}\caption{Numerical results for the the $S$-matrix element $\left|S_{32}\right|^{2}$ calculated for the different tilt profile tight-binding white hole analogues at a cut along one node ($p_2=0$ and $p_3=\pi/a$). We set $\alpha_{\rm t}=0.1$ for all tilt profiles. The results are compared to the theoretical function $X$ of Eq.~9 of the main text.\label{fig:DiffProfTransProb}}
\end{figure}

To confirm that the proposed analogue horizon physics is universal and dominated by the local rate of change of the tilt at the horizon, i.e. the analogue of the surface gravity, we check that other tilt profiles do not affect the exponential scaling of S-matrix elements. For this purpose, we choose two additional exemplary tilt profiles, which are different then the hyperbolic tangent profile $\ensuremath{V_{t,1}(x_{1})}$ of the main text, but have an approximately linear change at the horizon. The two tilt profiles are a linear profile, $\ensuremath{V_{lin,1}(x_{1})=\pm1\pm\alpha_{{\rm t}}\,x_{1}}$, and an arctangent profile, $\ensuremath{V_{atan,1}(x_{1})=\pm1\pm\frac{2}{\pi}arctan\left(\frac{\alpha_{{\rm t}}\,\pi\,x_{1}}{2}\right)}$. We calculate the transmission probability of the channel following $X$ (see $\left|S_{32}\right|^{2}$ in Fig.~(2) of the main text) for the three tilt profiles, while using the same rate $\alpha_{{\rm t}}$ for all profiles, and compare to the $X$ of Eq.~9 of the main text. We plot the results in Fig.~\ref{fig:DiffProfTransProb} and find that in the energy range $0-0.1$, the S matrix elements $\left|S_{32}\right|^{2}$ calculated with all profiles follow the theoretical curve, as expected. Beyond this range the results start to deviate from the theoretical curve to different extents. This indicates that experiments will be able to see the proposed physics, even if they do not realize a perfect tanh-profile.

\end{document}